\documentclass[showpacs,pra,twocolumn]{revtex4-1}

\usepackage{amsmath, amssymb, bm, graphicx, hyperref}
\usepackage{xcolor}
\usepackage[english]{babel}
\usepackage{physics}

\begin{document}
\title{Exact dynamics of first-order system--bath coherence in bilinear bosonic models}

\author{Yang-Yang Xie$^{1, 2, 9}$}

\author{Song-Lin Wu$^{3}$}

\author{Paul Brumer$^{4}$}

\author{Zhao-Ming Wang$^{1, 5, 6}$}
\email{wangzhaoming@ouc.edu.cn}

\author{Lian-Ao Wu$^{2, 7, 8}$}
\email{lianaowu@gmail.com}

\affiliation{
	$^{1}$ College of Physics and Optoelectronic Engineering, Ocean University of China, Qingdao 266100, China \\
	$^{2}$ Department of Physics, University of the Basque Country UPV/EHU, Bilbao 48080, Spain \\
	$^{3}$ School of Physics and Materials Engineering, Dalian Nationalities University, Dalian 116600, China \\
	$^{4}$ Chemical Physics Theory Group, Department of Chemistry, and Center for Quantum Information and Quantum Control, University of Toronto, Toronto, Ontario, Canada M5S 3H6 \\
	$^{5}$ Engineering Research Center of Advanced Marine Physical Instruments and Equipment of Ministry of Education, Ocean University of China, Qingdao 266100, China \\
	$^{6}$ Qingdao Key Laboratory of Optics and Optoelectronics, Qingdao 266100, China \\
	$^{7}$ IKERBASQUE Basque Foundation for Science, Bilbao 48013, Spain \\
	$^{8}$ EHU Quantum Center, University of the Basque Country UPV/EHU, Bilbao 48940, Spain \\
	$^{9}$ School of Physics and Astronomy, Shanghai Jiao Tong University, Shanghai 200240, China
}

\date{\today}

\begin{abstract}
We investigate the exact dynamics of first-order system--bath coherence induced by excitation exchange in bilinear bosonic models. By solving the linear Heisenberg equations, we obtain the exact evolution of system and bath operators and evaluate the first-order coherence between the system mode and the collective bath mode directly coupled to it.	We analyze how this coherence depends on initial occupations, coupling strength, spectral width, and detuning, showing that its buildup and oscillatory behavior are closely related to excitation exchange and reservoir memory. We further examine controlled coherence dynamics under leakage-elimination-operator inspired modulation of the system frequency. The results show that random modulation can suppress excitation leakage and maintain finite and relatively stable coherence fluctuations at long times. These results clarify the evolution and control of first-order system--bath coherence in settings where a localized bosonic mode is coupled to a structured bosonic reservoir.
\end{abstract}

\maketitle

\section{Introduction}

Solving the dynamics of open quantum systems is one of the most challenging topics in quantum physics due to the inevitable interactions between realistic physical systems and their environments~\cite{breuer2002theory}. In most treatments, this influence is described through the reduced state of the system~\cite{rivas2011open}, whose evolution captures environmentally induced effects including decoherence~\cite{zurek2003decoherence}, dissipation~\cite{weiss2012quantum}, and relaxation~\cite{leggett1987dynamics}. However, such a reduced description necessarily discards the correlations established between the system and the bath during the evolution~\cite{breuer2016colloquium}. Among these correlations, system--bath coherence~\cite{garraway1997nonperturbative, ishizaki2008nonperturbative, dijkstra2010nonmarkovian, tanimura2014reduced, dodin2022noise, paul2018shedding} is of particular importance, as it directly reflects the phase coherence associated with the interaction. Further, it is crucial in the establishment of ubiquitous systems in the non-equilibrium steady state. An exact characterization of such coherence is therefore desirable for a deeper understanding of open system dynamics.

Many studies have clarified important aspects of system--environment correlations from various perspectives. Some studies have characterized system--environment correlations through quantum-information-inspired measures, including mutual information~\cite{li2017production}, entanglement measures~\cite{hilt2009system}, and relative-entropy-based quantifiers~\cite{maziero2012genuine}. These measures provide useful tools for quantifying correlations beyond the reduced system state and have helped clarify the roles of such correlations in phenomena such as decoherence and the redistribution of initial correlations during the evolution. Others have shown that such hidden correlations can be witnessed indirectly from reduced system dynamics~\cite{laine2011witness, rossatto2011purity, gessner2014local}. For example, an increase in the trace distance between reduced states has been identified as a witness of initial system--environment correlations, even when access to the environment is unavailable~\cite{smirne2011experimental}. In addition, a number of works have examined bath-related coherence phenomena in particular open system models, such as bath-induced intra-system coherence~\cite{eastham2016bath}, coherence trapping~\cite{addis2014coherence}, and steady-state coherence~\cite{guarnieri2018steady}. While these works have provided valuable insight, less attention has been paid to the direct dynamics of coherence between a system and the environmental degrees of freedom, especially in structured reservoirs where memory effects play an important role.

Additional approximations are usually required to obtain closed equations for reduced system dynamics, most commonly the Born, Markov, or secular approximations. Exactly solvable models of open quantum systems are therefore particularly valuable, as they not only provide genuine insight into the underlying microscopic mechanisms, but also serve as benchmarks for various approximation schemes. Bilinear bosonic models have long been recognized as one such rare platform, for which exact treatments can be developed even for general spectral densities~\cite{hu1992quantum, karrlein1997exact, zhang2012general, homa2023analytical}. They describe a broad class of physical settings in which a localized bosonic mode is effectively linearly coupled to a continuum of environmental modes, such as a cavity mode coupled to photonic- or phononic-crystal reservoirs~\cite{coles2014waveguidecoupled, gomis-bresco2014a, ren2020two}. Recently, Wang \textit{et al.} further established an explicit, physically transparent, and analytically tractable framework for this class of models under Lorentzian spectral densities~\cite{wang2025analytical}. This advance makes it possible to track the system and bath operators directly, and has enabled subsequent extensions to two-mode settings, where the exact dynamics of coherence between system modes were investigated in both common and individual structured reservoirs~\cite{wu2025steady, abu-nada2026dynamics}. However, even within this exactly solvable framework, the coherence established between the system and the bath has not yet been systematically analyzed.

In this work, we study the exact dynamics of first-order system--bath coherence established between a system oscillator and the collective bath mode coupled to it in bilinear bosonic models. We evaluate this coherence directly from the exact Heisenberg operator evolution and analyze its dependence on the initial occupations, coupling strength, spectral width, and system--bath detuning. We further introduce leakage-elimination-operator (LEO)-inspired modulation of the system frequency and compare two representative modulation protocols to examine how external control modifies the coherence dynamics associated with system excitation leakage.

This paper is organized as follows. In Sec.~\ref{sec_model}, we present the Hamiltonian of bilinear bosonic models, derive the exact equations of motion for the system and bath operators, and obtain the analytical solutions for Lorentzian spectral densities. In Sec.~\ref{sec_coherence}, we introduce the first-order system--bath coherence and discuss its physical meaning in this context. In Sec.~\ref{sec_free}, we investigate the free coherence dynamics and their dependence on system and bath parameters. In Sec.~\ref{sec_controlled}, we examine the controlled coherence dynamics under LEO-inspired system frequency modulation. Finally, we summarize our main findings in Sec.~\ref{sec_con}.

\section{Model and exact dynamics} \label{sec_model}

We consider a bosonic system mode linearly coupled to a bath consisting of independent bosonic modes. The total Hamiltonian is
\begin{align}
	H = \omega_0 a^\dagger a + \sum_j \omega_j b_j^\dagger b_j + \sum_j \left( g_j a^\dagger b_j + g_j^* a b_j^\dagger \right).
	\label{ham}
\end{align}
Throughout this work, we set $\hbar=1$ and omit operator hats for simplicity. Here $\omega_0$ and $\omega_j$ denote the frequencies of the system mode and the $j$-th bath mode, respectively. $a, a^\dagger$ and $b_j, b_j^\dagger$ are the corresponding annihilation and creation operators. $g_j$ denotes the complex coupling coefficient between them.

The Heisenberg equations for the system and bath operators are
\begin{align}
	\dot{a}   &= -i\omega_0 a - i \sum_j g_j b_j, \label{a_motion} \\
	\dot{b}_j &= -i \omega_j b_j - i g_j^* a.     \label{bj_motion}
\end{align}
Due to the linear form of these equations, the time-evolved system operators can be expressed as a linear combination of the initial operators
\begin{align}
	a(t) = A(t)a(0) + \sum_{j}B_{j}(t)b_{j}(0).
	\label{a_expansion}
\end{align}
Here $A(t)$ and $B_{j}(t)$ are time-dependent expansion coefficients, which satisfy the initial conditions $A(0)=1$ and $B_j(0)=0$. To determine these coefficients, we first formally integrate Eq.~(\ref{bj_motion}) to obtain
\begin{align}
	b_j(t) = b_j(0) e^{-i \omega_j t} - i g_j^* \int_0^t ds \, a(s) e^{-i \omega_j (t - s)}.  
	\label{bj_integral}
\end{align}
Substituting the result into Eq.~(\ref{a_motion}) yields the integro-differential equation for the system operator
\begin{align}
	\dot{a} = - i\omega_0a(t) - \int_{0}^{t} ds \, G(t-s) a(s) + F(t),
	\label{a_integral_diff} 
\end{align}
with the inhomogeneous noise term $F(t) = - i\sum_{j}g_{j}b_{j}(0)e^{-i\omega_{j}t}$ and the memory kernel $G(t-s) = \sum_{j}|g_{j}|^{2} e^{-i\omega_j(t-s)} \to \int d\omega \, J(\omega) e^{-i\omega(t-s)}$ in the continuum limit. Substituting Eq.~(\ref{a_expansion}) into Eq.~(\ref{a_integral_diff}) gives the first-order equations for the coefficients
\begin{align}
	\dot{A}(t) & = -i\omega_0A(t)-\int_{0}^{t}ds \, G(t-s)A(s), 
	\label{dot_A} \\
	\dot{B}_{j}(t) & = -i\omega_0B_{j}(t) - \int_{0}^{t}ds \, G(t-s)B_{j}(s) - ig_{j}e^{-i\omega_{j}t}, 
	\label{dot_B}
\end{align}
with the initial conditions $\dot{A}(0) = -i\omega_0$ and $\dot{B}_j(0) = -ig_{j}$. Further differentiation of Eqs.~(\ref{dot_A}) and (\ref{dot_B}) gives the corresponding second-order equations
\begin{align}
	\ddot{A}(t)  
	= & - i\omega_0\dot{A}(t)-G(0)A(t) - \int_{0}^{t} ds \, \dot{G}(t-s)A(s), 	\label{ddot_A} \\
	\ddot{B}_{j}(t) 
	= & - i\omega_0\dot{B}_{j}(t) - G(0)B_{j}(t) - \int_{0}^{t} ds \, \dot{G}(t-s)B_{j}(s) \notag \\
	& - g_{j}\omega_{j}e^{-i\omega_{j}t}. 
	\label{ddot_B}
\end{align}
Note that Eqs.~(\ref{ddot_A}) and (\ref{ddot_B}), together with their initial conditions, determine the exact dynamics of the system operator $a(t)$ within the model~(\ref{ham}). No approximations, such as the Born or Markov approximations, are introduced in this derivation. The Hamiltonian~(\ref{ham}) contains only rotating-wave terms. If counter-rotating interactions are retained, the operator expansion must be enlarged to include creation operators.

Moreover, Eqs.~(\ref{ddot_A}) and (\ref{ddot_B}) become analytically tractable when the memory kernel can be written as a finite sum of exponentials. In this work, we choose the Lorentzian spectral density
\begin{align}
	J(\omega) = \frac{\Gamma}{2\pi}\frac{\gamma^{2}}{(\omega-\Omega)^{2}+\gamma^{2}},
	\label{J}
\end{align}
which gives the exponentially decaying memory kernel
\begin{align}
	G(t-s) = \frac{\Gamma\gamma}{2} e^{-(\gamma+i\Omega)|t-s|},
\end{align}
and therefore
\begin{align}
	\dot{G}(t-s)=-(\gamma+i\Omega)G(t-s).
\end{align}
Here $\Gamma$ describes the overall system--bath coupling strength, $\gamma$ characterizes the spectral width, and $\Omega$ denotes the central frequency of the spectral density.

With this spectral density, Eqs.~(\ref{ddot_A}) and (\ref{ddot_B}) reduce to the second-order linear ordinary differential equations
\begin{align}
	\ddot{A}(t) + c_1 \dot{A}(t) + c_2 A(t) &= 0,            	\label{ode_A} \\
	\ddot{B}_{j}(t) + c_1 \dot{B}_j(t) + c_2 B_j(t) &= d_j(t), 	\label{ode_B}
\end{align}
with the coefficients $c_1=\gamma+i(\omega_0+\Omega)$, $c_2=\Gamma\gamma/2-\omega_0\Omega+i\gamma\omega_0$ and the driving term $d_j(t) = -g_j[(\omega_j-\Omega)+i\gamma]e^{-i\omega_j t}$.

The solution of Eq.~(\ref{ode_A}) is
\begin{align}
	A(t) = A_+ e^{\lambda_+ t} + A_- e^{\lambda_- t}, 
	\label{sol_A}
\end{align}
where $A_\pm$ are determined by the initial conditions, and $\lambda_\pm = (-c_1\pm\sqrt{c_1^2-4c_2})/2$ are the roots of $\lambda^2 + c_1 \lambda + c_2 = 0$. The real and imaginary parts of $\lambda_\pm$ determine the decay and oscillatory behavior of $A(t)$, respectively. Eq.~(\ref{ode_B}) has the same homogeneous part as Eq.~(\ref{ode_A}) but contains an inhomogeneous term. Its solution can therefore be written as
\begin{align}
	B_j(t) = B_{j,+}e^{\lambda_+ t} + B_{j,-}e^{\lambda_- t} + B_{j,\mathrm{p}}(t).
	\label{sol_B}
\end{align}
Here the constants $B_{j,\pm}$ are fixed by the initial conditions $B_j(0)=0$ and $\dot B_j(0)=-ig_j$, and the particular solution
\begin{align}
	B_{j,\mathrm{p}}(t) = - \frac{g_{j}(\omega_{j}-\Omega+i\gamma)}{\Gamma\gamma/2-(\omega_{0}-\omega_{j})^{2}+i\gamma(\omega_{0}-\omega_{j})} e^{-i\omega_{j}t},
\end{align}
which can be obtained by substituting the ansatz $f_{j}e^{-i\omega_{j}t}$ into Eq.~(\ref{ode_B}). With the solutions (\ref{sol_A}) and (\ref{sol_B}), the system operator at time $t$ is fully determined.

The bath operator $b_j$ can be obtained by substituting the expansion of $a(s)$ into Eq.~(\ref{bj_integral}). This gives
\begin{align}
	b_j(t)
	&= - i g_j^{*}\int_0^t ds \, A(s) e^{-i\omega_j(t-s)} a(0) \notag \\
	&\quad - i g_j^{*}\sum_k \int_0^t ds \, B_k(s) e^{-i\omega_j(t-s)} b_k(0) + e^{-i\omega_j t} b_j(0) \notag \\
	&\equiv P_j(t) a(0) + \sum_k Q_{j,k}(t) b_k(0),
	\label{bj_expansion}
\end{align}
where 
\begin{align}
	P_j(t)     & = - i g_j^{*}\int_0^t ds \, A(s) e^{-i\omega_j(t-s)}, \\
	Q_{j,k}(t) & = - i g_j^{*}\int_0^t ds \, B_k(s) e^{-i\omega_j(t-s)} + \delta_{j,k} e^{-i\omega_j t}.
\end{align}

Since both system and bath operators evolve as linear combinations of all initial operators, the expectation values of system observables at time $t$ depend not only on the intrinsic system dynamics but also on the system--bath coherence generated during the evolution. This dependence underscores the necessity of quantifying such coherence, prompting us to introduce an operator-based system--bath coherence characterization in the following section.

\section{First-order system--bath coherence} \label{sec_coherence}

In this work, we focus on the first-order coherence between the system oscillator and the collective bath mode. We characterize this coherence by
\begin{align}
	C(t) = \langle a^\dagger(t)\mathcal{B}(t)\rangle,
	\label{c_propose}
\end{align}
which is defined in analogy with the standard notion of first-order coherence in quantum optics~\cite{mandel1995optical}. Here
\begin{align}
	\mathcal{B}(t)=\frac1{\mathcal{N}}\sum_j g_j b_j(t), \quad \mathcal{N} = \sqrt{\sum_j |g_j|^2}, 
\end{align} 
denotes the collective bath operator directly coupled to the system. This quantity can also be viewed as extending the notion of intermode coherence in the Schwinger boson representation to the system--bath context~\cite{schwinger1965on, mandel1995optical}.

Each term in Eq.~(\ref{c_propose}) can be written as
\begin{align}
	\langle a^\dagger(t)b_j(t)\rangle
	&=\mathrm{Tr}\!\left[ U^\dagger a^\dagger(0) U U^\dagger b_j(0) U \rho(0)\right]  \notag \\
	&=\mathrm{Tr}\!\left[ a^\dagger(0) b_j(0) \rho(t)\right]  \notag \\
	&=\sum_{\{n\}}\sqrt{(n_\mathrm{s}+1)n_j} \, \rho_{\{n\}; \{n_\mathrm{s}+1, n_j-1, \ldots\}}(t),
	\label{c_each}
\end{align}
where $U=\exp(-iHt)$ is generated by the Hamiltonian in Eq.~(\ref{ham}), $\rho(t)=U\rho(0)U^\dagger$, and $\{n\}$ denotes the joint occupation numbers in the Fock basis. Eq.~(\ref{c_each}) shows that $\langle a^\dagger(t)b_j(t)\rangle$ selects the off-diagonal elements of the joint system--bath density matrix $\rho(t)$ associated with the transfer of one excitation from the $j$-th bath mode to the system mode, with all other occupations unchanged.

With the decompositions of $a(t)$ and $b_j(t)$ in Eqs.~(\ref{a_expansion}) and (\ref{bj_expansion}), we can easily derive the coherence
\begin{align}
	C(t) = \frac{1}{\mathcal N}\sum_j g_j [ A^*(t)P_j(t) n_\mathrm{s} + \sum_k B_k^*(t)Q_{j,k}(t) n_k ]. 
	\label{c_multimode}
\end{align}
where $n_\mathrm{s}$ and $n_k$ are the initial excitation numbers of the system and $k$-th bath. We consider that the system and bath are initialized to their respective thermal states and the bath are at the same temperature $T_\mathrm{b}$, so that $n_\mathrm{s} = \langle a^\dagger(0) a(0)\rangle=[\exp(\omega_0/T_\mathrm{s})-1]^{-1}$, $n_k=\langle b_k^\dagger(0) b_k(0)\rangle=[\exp(\omega_k/T_\mathrm{b})-1]^{-1}$. Eq.~(\ref{c_multimode}) naturally decomposes into system-induced and bath-induced contributions, which we denote by $C_\mathrm{s}(t)$ and $C_\mathrm{b}(t)$. In the continuum limit of the bath modes, the discrete sums are replaced by integrals $\int d\omega$ weighted by the appropriate spectral weight.

It is worth noting that, for nonclassical initial states, the bilinear interaction may induce entanglement through the coherent superposition of indistinguishable bosonic paths, but this type of entanglement is qualitatively different from that produced by state-dependent conditional dynamics, such as those enabled by nonlinear interactions, which provide the entangling operations required for universal quantum computation. A brief illustration of the two types of entanglement is given in the Appendix.

\section{Free dynamics} \label{sec_free}

In this section, we investigate how the first-order system--bath coherence amplitude $|C|$ depends on the system and bath parameters, namely the system temperature $T_\mathrm{s}$, bath temperature $T_\mathrm{b}$, coupling strength $\Gamma$, spectral width $\gamma$, and the detuning between the system frequency and the bath central frequency $\Delta=\Omega-\omega_0$. It is clear from Eq.~(\ref{c_multimode}) that $T_\mathrm{s}$ and $T_\mathrm{b}$ enter $C$ explicitly through the initial occupations, whereas $\Gamma$, $\gamma$, and $\Delta$ primarily affect the dynamical coefficients $A$, $B_j$, $P_j$, and $Q_{j,k}$ governing the excitation exchange process. In what follows, we also present the corresponding dependence of the sytem occupation number $n_\mathrm{s}$ on these parameters, which provides a useful reference for understanding the resulting behavior of the coherence amplitude.

\begin{figure}[!tb]
	\centering
	\includegraphics[width=\columnwidth]{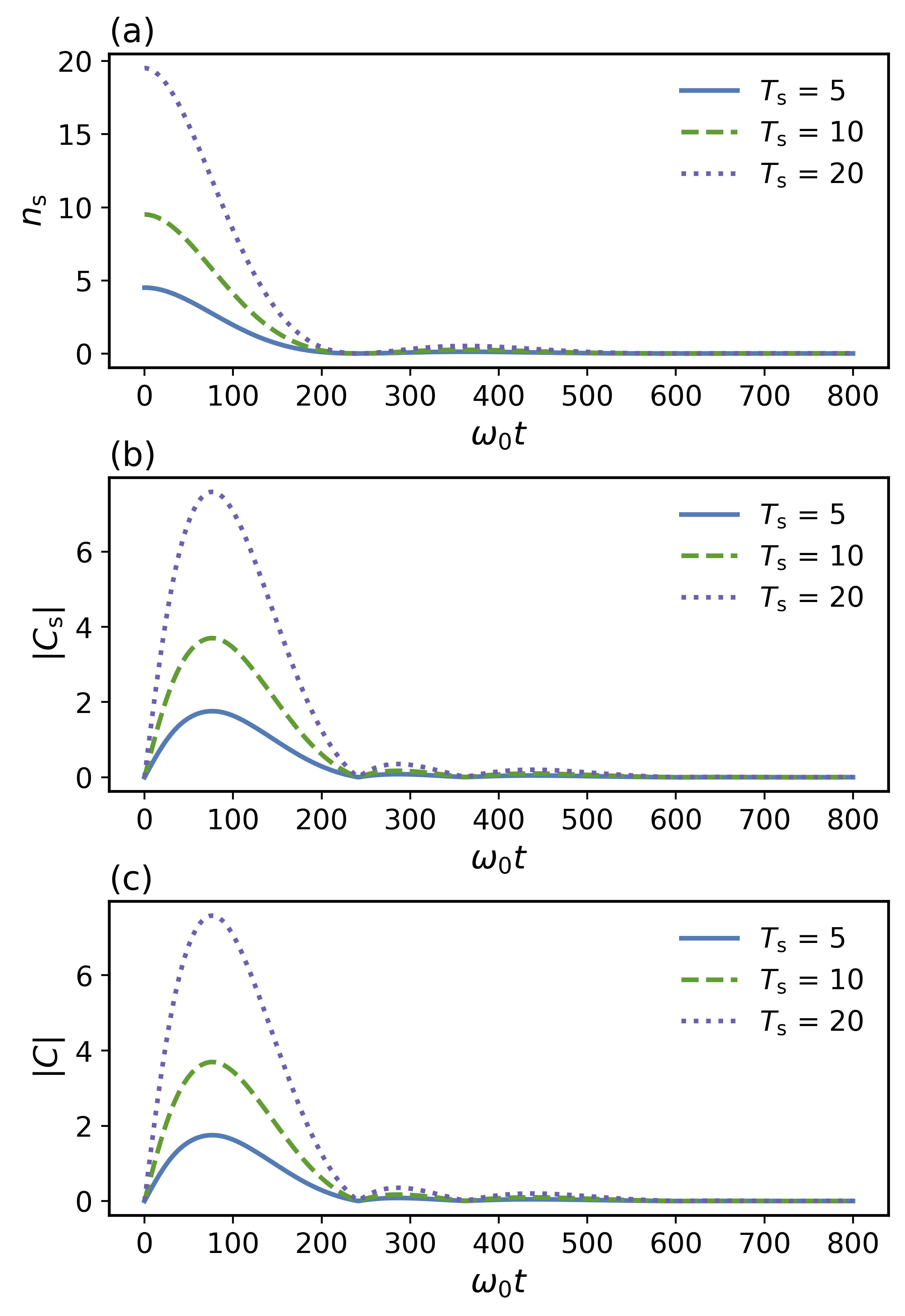}
	\caption{Effect of the system temperature $T_\mathrm{s}$ on the first-order system--bath coherence dynamics. (a) System occupation number $n_\mathrm{s}$. (b) System-induced coherence contribution $|C_\mathrm{s}|$. (c) Total first-order system--bath coherence amplitude $|C|=|C_\mathrm{s}+C_\mathrm{b}|$. Other parameters are $\omega_0=\Omega=1$, $\Gamma=0.02$, $\gamma=0.01$, and $T_\mathrm{b}=10$.}
	\label{fig1}
\end{figure}
We first focus on the system temperature $T_\mathrm{s}$, which enters Eq.~(\ref{c_multimode}) only through the initial system excitation number $n_\mathrm{s}$. Fig.~\ref{fig1} plots the time evolution of the system occupation number $n_\mathrm{s}$, the system-induced coherence contribution $|C_\mathrm{s}|$, and the total coherence amplitude $|C|=|C_\mathrm{s}+C_\mathrm{b}|$ for different $T_\mathrm{s}$, with the other parameters fixed at $\omega_0=\Omega=1$, $\Gamma=0.02$, $\gamma=0.01$, and $T_\mathrm{b}=1$. Since $n_\mathrm{s} = [\exp(\omega_0/T_\mathrm{s})-1]^{-1}$, increasing $T_\mathrm{s}$ raises the initial system occupation and hence the initial system energy, as shown in Fig.~\ref{fig1}(a). This leads to a larger net energy transfer from the system to the bath during the subsequent evolution, accompanied by a larger system-induced contribution $|C_\mathrm{s}|$, whose magnitude increases markedly with $T_\mathrm{s}$ in Fig.~\ref{fig1}(b). By contrast, the bath-induced contribution $|C_\mathrm{b}|$ is independent of $T_\mathrm{s}$, and it remains much smaller than $|C_\mathrm{s}|$. Consequently, the total coherence amplitude in Fig.~\ref{fig1}(c) is dominated by the system-induced contribution and follows essentially the same temperature dependence: increasing the system temperature mainly enhances the transient coherence amplitude, while leaving its oscillatory behavior and long-time value unaffected.

It is worth noting that $|C(t)|$ does not decay to zero at long times, but retains a small residual amplitude. In the long-time limit, Eqs.~(\ref{sol_A}) and (\ref{sol_B}) show that the coefficient $A(t)$ vanishes, while $B_j(t)$ approaches its particular solution $B_{j,\mathrm{p}}(t)$. Since $B_{j,\mathrm{p}}(t)$ is driven by the driving term $d_j(t)$ and still carries the phase factor $e^{-i\omega_j t}$, the bath-related part of Eq.~(\ref{c_multimode}) can retain weak oscillatory components. Therefore, weak residual first-order coherence between the system and the bath can remain at long times~\cite{dodin2021generalized}.

\begin{figure}[!tb]
	\centering
	\includegraphics[width=\columnwidth]{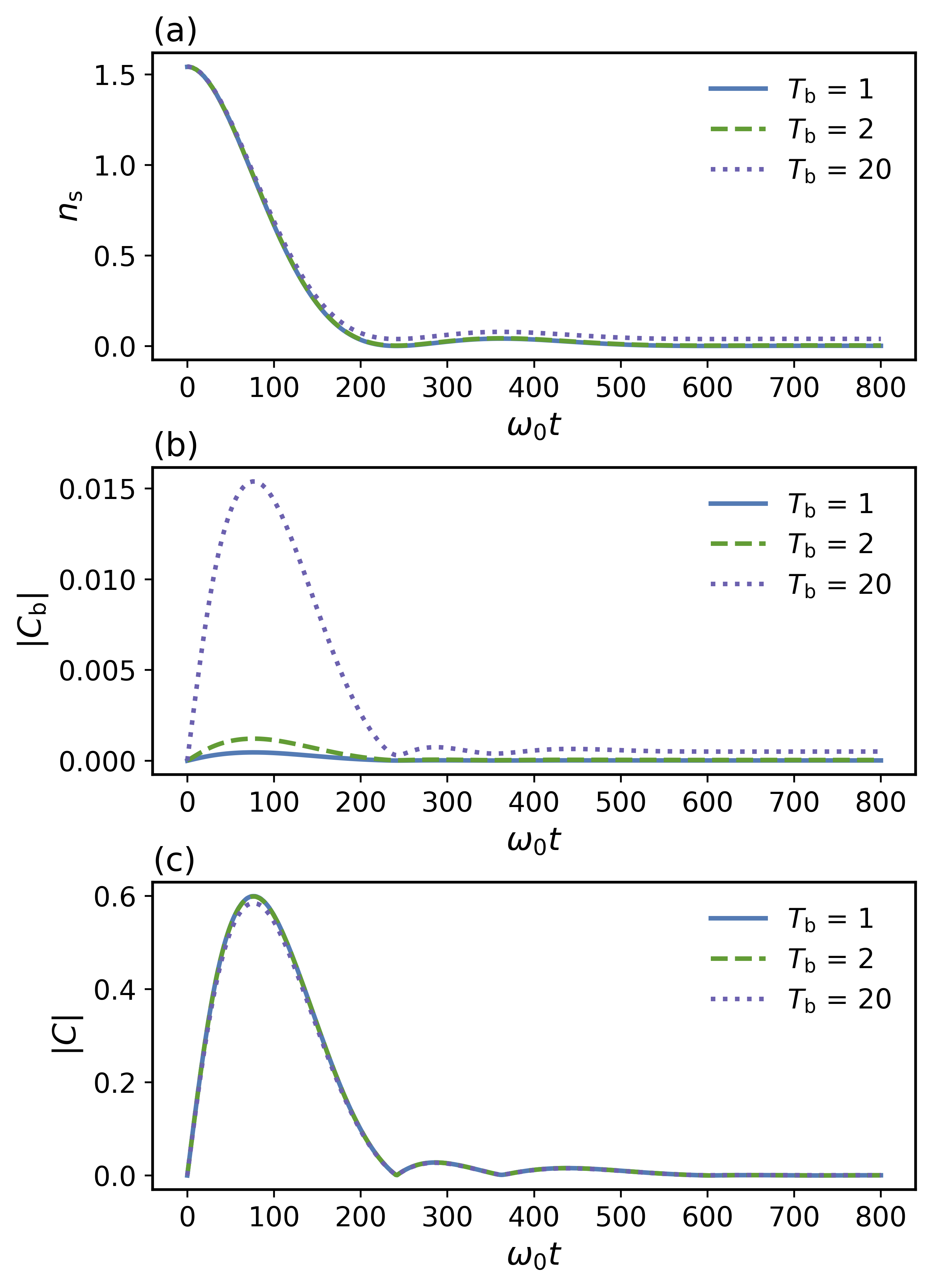}
	\caption{Effect of the bath temperature $T_\mathrm{b}$ on the first-order system--bath coherence dynamics. (a) System occupation number $n_\mathrm{s}$. (b) Bath-induced coherence contribution $|C_\mathrm{b}|$. (c) Total first-order system--bath coherence amplitude $|C|=|C_\mathrm{s}+C_\mathrm{b}|$. Other parameters are $\omega_0=\Omega=1$, $\Gamma=0.02$, $\gamma=0.01$, and $T_\mathrm{s}=2$.}
	\label{fig2}
\end{figure}
We now turn to the bath temperature $T_\mathrm{b}$, which enters Eq.~(\ref{c_multimode}) through the initial bath occupations $n_k = [\exp(\omega_k/T_\mathrm{b})-1]^{-1}$. Fig.~\ref{fig2} shows the system occupation number $n_\mathrm{s}$, bath-induced coherence contribution $|C_\mathrm{b}|$, and the total coherence amplitude $|C|=|C_\mathrm{s}+C_\mathrm{b}|$ for different $T_\mathrm{b}$, with $\omega_0=\Omega=1$, $\Gamma=0.02$, $\gamma=0.01$, and $T_\mathrm{s}=1$. As shown in Fig.~\ref{fig2}(a), a hotter bath enhances thermal backflow to the system, thereby reducing the net energy released from the system into the bath. This effect becomes more evident in the long-time limit, where the system occupation number $n_\mathrm{s}$ approaches a higher value for larger $T_\mathrm{b}$. In Fig.~\ref{fig2}(b), increasing $T_\mathrm{b}$ raises the initial bath occupations $n_k$, and therefore enhances the bath-induced contribution $|C_\mathrm{b}|$, while it remains much smaller than $|C_\mathrm{s}|$ in the parameter regime. As a result, the total coherence amplitude is still dominated by the system-induced contribution. Varying the bath temperature only weakly modifies its magnitude, including its long-time value, while leaving the overall temporal behavior nearly unchanged.

Before investigating the effects of the coupling strength $\Gamma$ and the spectral width $\gamma$, it is useful to recall how these two parameters enter the exact dynamics. For a Lorentzian spectral density, the expansion coefficients $A(t)$ and $B_j(t)$ are determined by the characteristic roots of Eqs.~(\ref{ode_A}) and (\ref{ode_B}), whose damping character depends on the relative magnitude of $\Gamma$ and $\gamma$. In particular, in the non-Markovian regime, the dynamics are underdamped for $\gamma < 2\Gamma$, leading to oscillatory behavior. For $\gamma > 2\Gamma$, the dynamics become overdamped and the decay is essentially monotonic~\cite{wang2025analytical}.

\begin{figure}[!tb]
	\centering
	\includegraphics[width=\columnwidth]{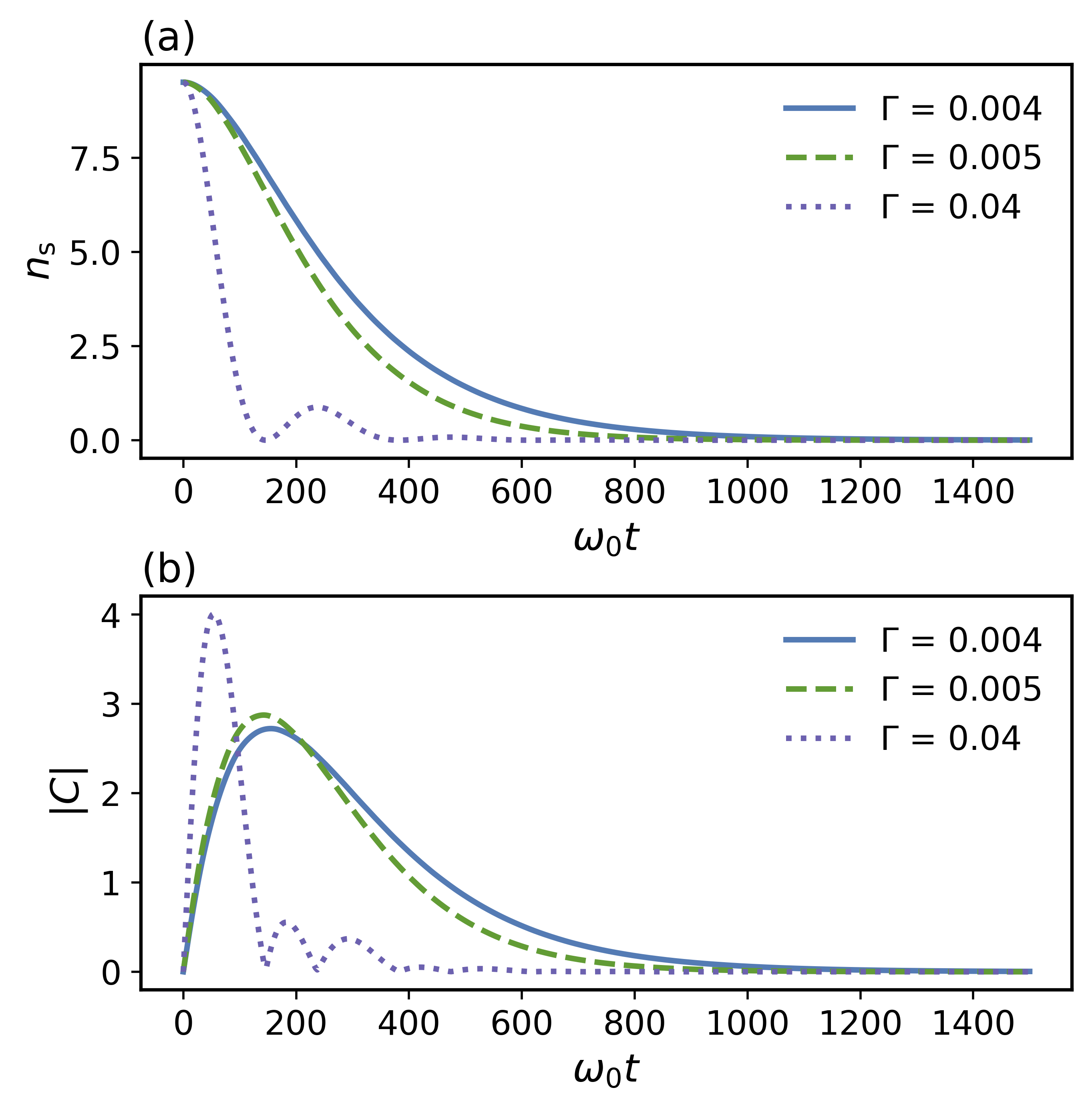}
	\caption{Effect of the coupling strength $\Gamma$ on the first-order system--bath coherence dynamics. (a) System occupation number $n_\mathrm{s}$. (b) First-order system--bath coherence amplitude $|C|$. Other parameters are $\omega_0=\Omega=1$, $\gamma=0.01$, $T_\mathrm{s}=10$, and $T_\mathrm{b}=1$.}
	\label{fig3}
\end{figure}
In Fig.~\ref{fig3}, we plot the time evolution of $n_\mathrm{s}$ and $|C|$ for different coupling strengths $\Gamma$, with $\omega_0=\Omega=1$, $\gamma=0.01$, $T_\mathrm{s}=10$, and $T_\mathrm{b}=1$. The values of $\Gamma$ correspond respectively to the overdamped ($\gamma>2\Gamma$), critically damped ($\gamma=2\Gamma$), and underdamped ($\gamma<2\Gamma$) regimes. In the overdamped and critically damped cases, the system occupation number $n_\mathrm{s}(t)$ decays smoothly, while the total coherence amplitude $|C(t)|$ exhibits a single broad peak before relaxation. As the coupling strength $\Gamma$ increases, $n_\mathrm{s}(t)$ decays more rapidly into the bath, causing $|C(t)|$ to rise more quickly to a higher peak. In the underdamped case, both $n_\mathrm{s}(t)$ and $|C(t)|$ exhibit pronounced oscillations, reflecting repeated energy exchange between the system and the bath. Overall, increasing the system--bath coupling strength accelerates the excitation leakage and changes the damping character of the total coherence amplitude from overdamped or critically damped relaxation to underdamped oscillatory behavior. It also slightly increases the long-time coherence amplitude, although this effect is weak on the scale of Fig.~\ref{fig3}.

Before discussing the role of the spectral width $\gamma$, recall that, for the Lorentzian spectral density, $\gamma^{-1}$ sets the memory time. A smaller $\gamma$ therefore corresponds to stronger memory effects. In the opposite limit $\gamma\to\infty$, the spectral density becomes flat, $J(\omega) \to \frac{\Gamma}{2\pi}$, and the memory kernel becomes effectively local in time, $G(t-s) \to \Gamma\delta(t-s)$, recovering the Markovian limit. The first-order equations for the coefficients $A(t)$ and $B_j(t)$, i.e., Eqs.~(\ref{ode_A}) and (\ref{ode_B}), reduce to
\begin{align}
	\dot A(t)   &= -i\omega_0A(t) - \frac\Gamma2 A(t), \\
	\dot B_j(t) &= -\left( \frac\Gamma2+i\omega_0 \right) B_j(t) - ig_j e^{-i\omega_j t}.
\end{align}
Together with the initial conditions $A(0)=1$ and $B_j(0)=0$, the equations yield
\begin{align}
	A(t)   &= e^{-(\Gamma/2+i\omega_0)t}, \\
	B_j(t) &= \frac{g_j}{-(\omega_0-\omega_j) + i\Gamma/2} \left(e^{-i\omega_jt}-e^{-(i\omega_0+\Gamma/2)t}\right).
\end{align}
Although the analytical expression for the first-order system--bath coherence $C(t)$ at finite times remains cumbersome even in this limit, its behavior can be readily evaluated numerically.

\begin{figure}[!tb]
	\centering
	\includegraphics[width=\columnwidth]{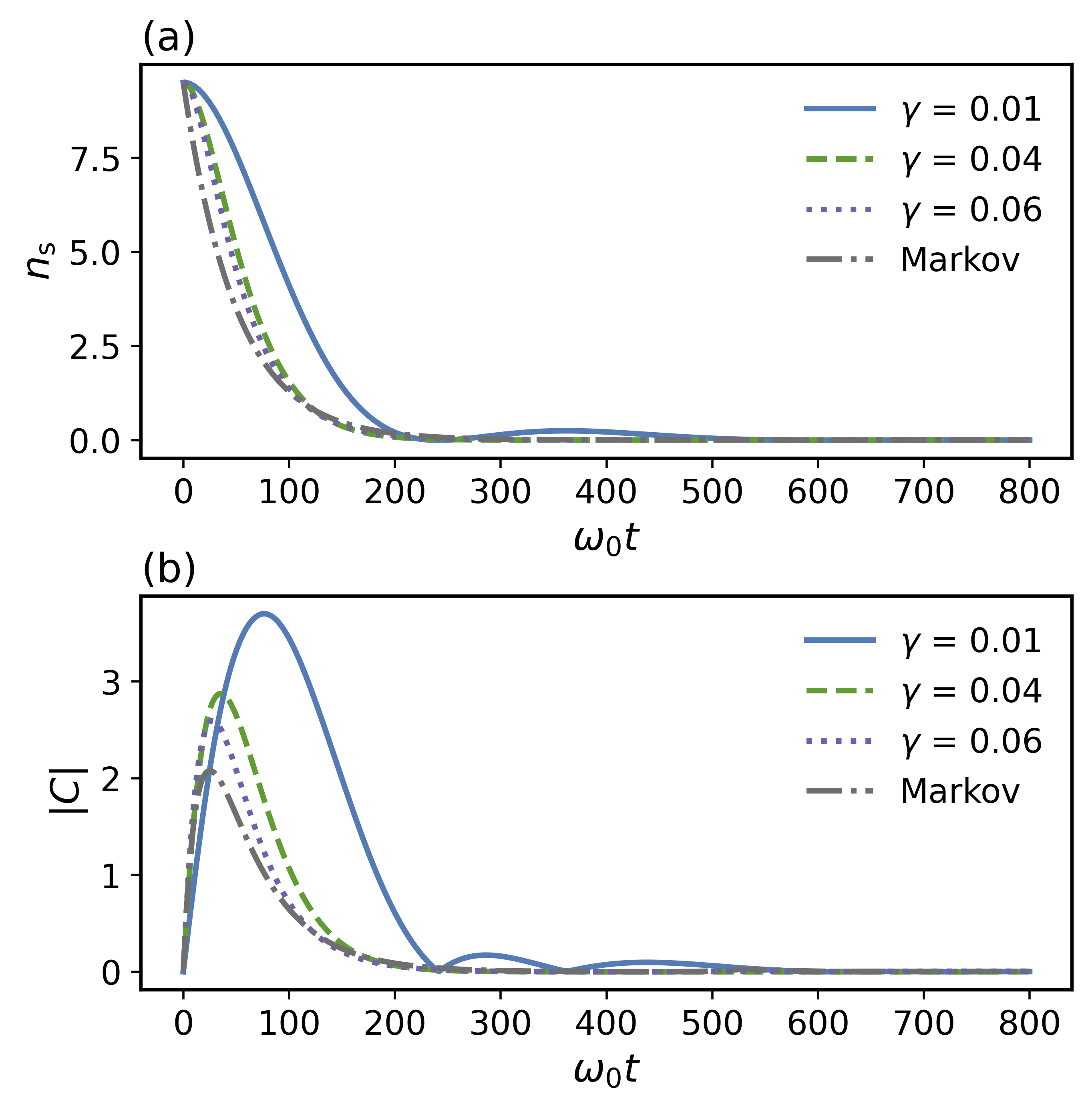}
	\caption{Effect of the spectral width $\gamma$ on the first-order system--bath coherence dynamics. (a) System occupation number $n_\mathrm{s}$. (b) First-order system--bath coherence amplitude $|C|$. Other parameters are $\omega_0=\Omega=1$, $\Gamma=0.02$, $T_\mathrm{s}=10$, and $T_\mathrm{b}=1$.}
	\label{fig4}
\end{figure}
In Fig.~\ref{fig4}, we show $n_\mathrm{s}(t)$ and $|C(t)|$ for various spectral widths $\gamma$, with $\omega_0=\Omega=1$, $\Gamma=0.02$, $T_\mathrm{s}=10$, and $T_\mathrm{b}=1$. Since $\gamma^{-1}$ sets the bath memory time, decreasing $\gamma$ allows the bath to retain information about the system for a longer duration. This is reflected by the slower relaxation of the system occupation number $n_\mathrm{s}(t)$ and the delayed buildup of the total coherence amplitude $|C(t)|$. The longer memory also enhances the transient excitation exchange between the system and the bath, leading to a larger peak and more visible oscillatory features of $|C(t)|$. In contrast, in the Markovian limit, the system energy decays more rapidly into the bath, corresponding to a fast buildup and subsequent decay of $|C(t)|$. In short, decreasing the spectral width delays the buildup of the coherence amplitude while enhancing its peak, whereas the Markovian limit exhibits the fastest buildup and the smallest peak. The coherence amplitude at long times also shows a slight dependence on the spectral width, although this effect is barely visible in Fig.~\ref{fig4}.

\begin{figure}[!tb]
	\centering
	\includegraphics[width=\columnwidth]{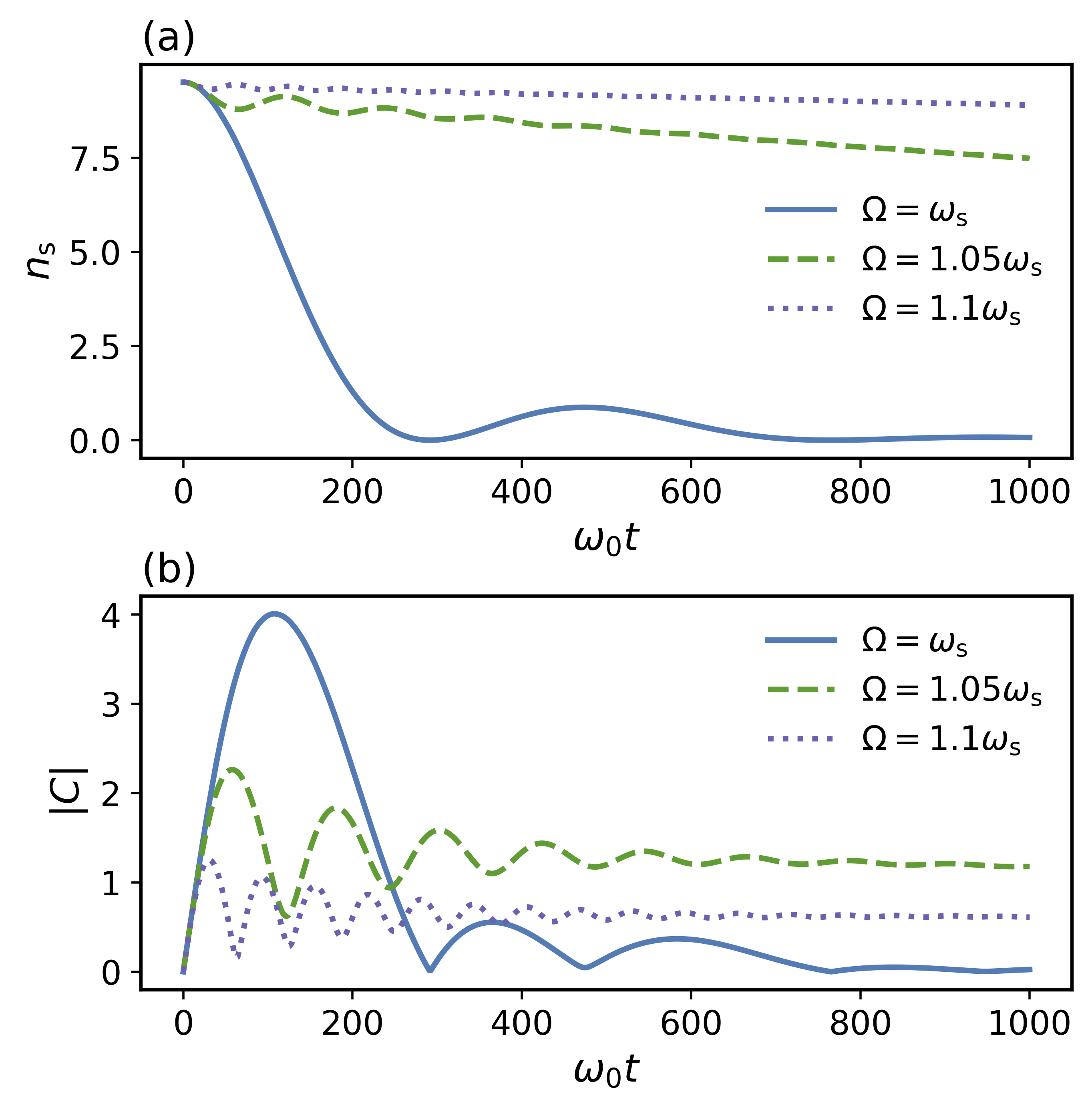}
	\caption{Effect of the detuning $\Delta = \Omega - \omega_0$ on the first-order system--bath coherence dynamics. (a) System occupation number $n_\mathrm{s}$. (b) First-order system--bath coherence amplitude $|C|$. Other parameters are $\omega_0=1$, $\Gamma=0.02$, $\gamma=0.005$, $T_\mathrm{s}=10$, and $T_\mathrm{b}=1$.}
	\label{fig5}
\end{figure}
We finally examine the effect of detuning between the system frequency and the bath central frequency, $\Delta = \Omega - \omega_0$. Since the Markovian limit corresponds to a flat spectral density independent of $\Omega$, detuning with respect to $\Omega$ is relevant only for non-Markovian dynamics. Fig.~\ref{fig5} shows $n_\mathrm{s}(t)$ and $|C(t)|$ for different detunings $\Delta$ in the non-Markovian regime, with $\omega_0=1$, $\Gamma=0.02$, $\gamma=0.005$, $T_\mathrm{s}=10$, and $T_\mathrm{b}=1$. In the resonant case, the frequency matching between the system and the dominant bath modes leads to pronounced energy transfer from the system into the bath, as reflected by the rapid decay of the system occupation number $n_\mathrm{s}(t)$ and the large peak of the total coherence amplitude $|C(t)|$. When detuning is introduced, the frequency mismatch weakens the system--bath excitation exchange. As a result, a larger fraction of $n_\mathrm{s}$ remains localized in the system, while the peak amplitude of $|C(t)|$ is reduced. At the same time, the detuned cases exhibit more frequent oscillations in $|C(t)|$, as the frequency mismatch introduces an additional phase-evolution timescale. Overall, detuning suppresses the buildup of a large coherence amplitude while introducing faster phase-induced oscillations in its dynamics.

The results above characterize the free evolution of the first system--bath coherence and show that its behavior is closely related to the excitation exchange betweeon the system oscillator and the structured bath. In particular, the coherence buildup and decay are strongly affected by the system--bath detuning. This observation suggests that external modulation of the system frequency may provide an effective way to modify the exchange process and thereby control the coherence dynamics. We now examine such controlled dynamics in the following section.

\section{Controlled dynamics under system frequency modulation} \label{sec_controlled}

We introduce a LEO-inspired frequency modulation~\cite{wu2002efficient, wu2009master, jing2014one}. The basic idea of LEO control is to apply fast controls that average unwanted couplings nearly to zero over time. LEO theory also leads to a counterintuitive observation: random pulses can still suppress unwanted couplings, provided that their characteristic frequency is sufficiently higher than that of the unwanted couplings. In the present model, we implement the control by allowing the system frequency to become time dependent,
\begin{align}
	\omega_0 \rightarrow \omega_\mathrm{s}(t) = \omega_0 + \delta\omega(t),
\end{align}
where $\omega_0$ is the bare system frequency and $\delta\omega(t)$ is an externally applied frequency shift. The operator-based coherence definition remains unchanged. However, the coherence dynamics are evaluated from the first-order equations, Eqs.~(\ref{dot_A}) and (\ref{dot_B}), with the time-dependent $\omega_\mathrm{s}(t)$, because the corresponding second-order form would involve $\dot{\omega}_\mathrm{s}(t)$.

We consider two representative modulation protocols: rectangular modulation and random modulation. For the rectangular modulation, 
	\begin{align}
		\delta\omega(t) =
		\begin{cases}
			\delta\omega_\mathrm{max}, \, & t\in[n\tau_\mathrm{rect},(n+1)\tau_\mathrm{rect}), \, n~\text{is even}, \\
			0, \, & t\in[n\tau_\mathrm{rect},(n+1)\tau_\mathrm{rect}), \, n~\text{is odd},
		\end{cases}
	\end{align}
	so that the frequency shift alternates between $\delta\omega_\mathrm{max}$ and 0 after each interval $\tau_\mathrm{rect}$. For the random modulation, 
	\begin{align}
		\delta\omega(t) = \delta\omega_n, \quad t\in[n\tau_\mathrm{ran},(n+1)\tau_\mathrm{ran}),
	\end{align}
where the random shifts $\delta\omega_n$ are independently sampled from a uniform distribution over $[0, \delta\omega_\mathrm{max}]$. Here $\delta\omega_\mathrm{max}$ denotes the maximum frequency shift, while $\tau_\mathrm{rect}$ and $\tau_\mathrm{ran}$ set the switching intervals of the rectangular and random modulations, respectively. In what follows, we examine the effects of the two modulation protocols on the first-order system--bath coherence, with the open-system parameters fixed at $\omega_0=\Omega=1$, $\Gamma=0.02$, $\gamma=0.01$, $T_\mathrm{s}=10$, and $T_\mathrm{b}=1$.

\begin{figure}[!tb]
	\centering
	\includegraphics[width=\columnwidth]{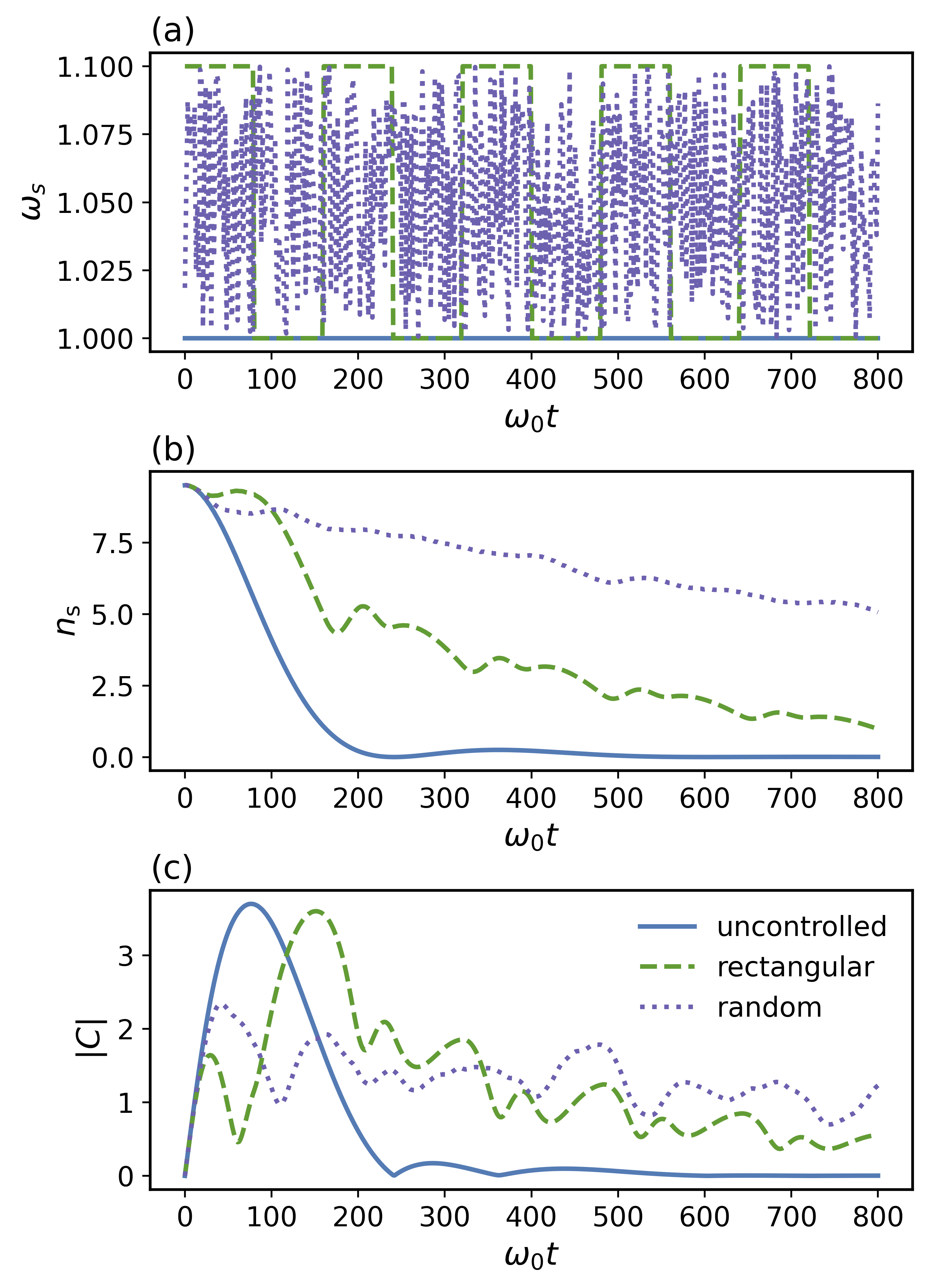}
	\caption{Controlled dynamics under rectangular and random modulations of the system frequency. (a) System frequency $\omega_\mathrm{s}$. (b) System occupation number $n_\mathrm{s}$. (c) First-order system–bath coherence amplitude $|C|$. The modulation parameters are $\delta\omega_\mathrm{max}=0.1$, $\omega_0\tau_\mathrm{rect}=80$ and $\omega_0\tau_\mathrm{ran}=1.6$.}
	\label{fig6}
\end{figure}
We first present the modulated system frequency $\omega_\mathrm{s}(t)$ in Fig.~\ref{fig6}(a), and the controlled dynamics of the system occupation number $n_\mathrm{s}(t)$ and the coherence amplitude $|C(t)|$ in Figs.~\ref{fig6}(b) and \ref{fig6}(c), respectively. The modulation parameters $\delta\omega_\mathrm{max}=0.1$, $\omega_0\tau_\mathrm{rect}=80$ and $\omega_0\tau_\mathrm{ran}=1.6$. In Fig.~\ref{fig6}(b), the uncontrolled system excitation almost completely leaks into the bath, whereas the rectangular modulation delays the leakage but still leads to a substantial decrease. By contrast, random modulation gives the strongest protection, keeping $n_\mathrm{s}(t)$ at a higher level throughout the evolution. The coherence dynamics in Fig.~\ref{fig6}(c) show a similar contrast. Without control, $|C(t)|$ rapidly builds up to a larger peak and then decays to nearly zero. Under rectangular modulation, the peak is delayed and strong oscillatory behavior appears, but the overall coherence amplitude still decreases significantly. Under random modulation, the peak is the smallest, but $|C(t)|$ continues to fluctuate around a higher long-time level than in the rectangular case. In short, although the rectangular modulation can slow down the system-excitation dissipation and coherence decay, its periodic structure still allows recurrent excitation exchange between the system and the bath; random modulation more effectively prevents the complete decay of the first-order system--bath coherence and maintains coherence fluctuations at long times.

\begin{figure}[!tb]
	\centering
	\includegraphics[width=\columnwidth]{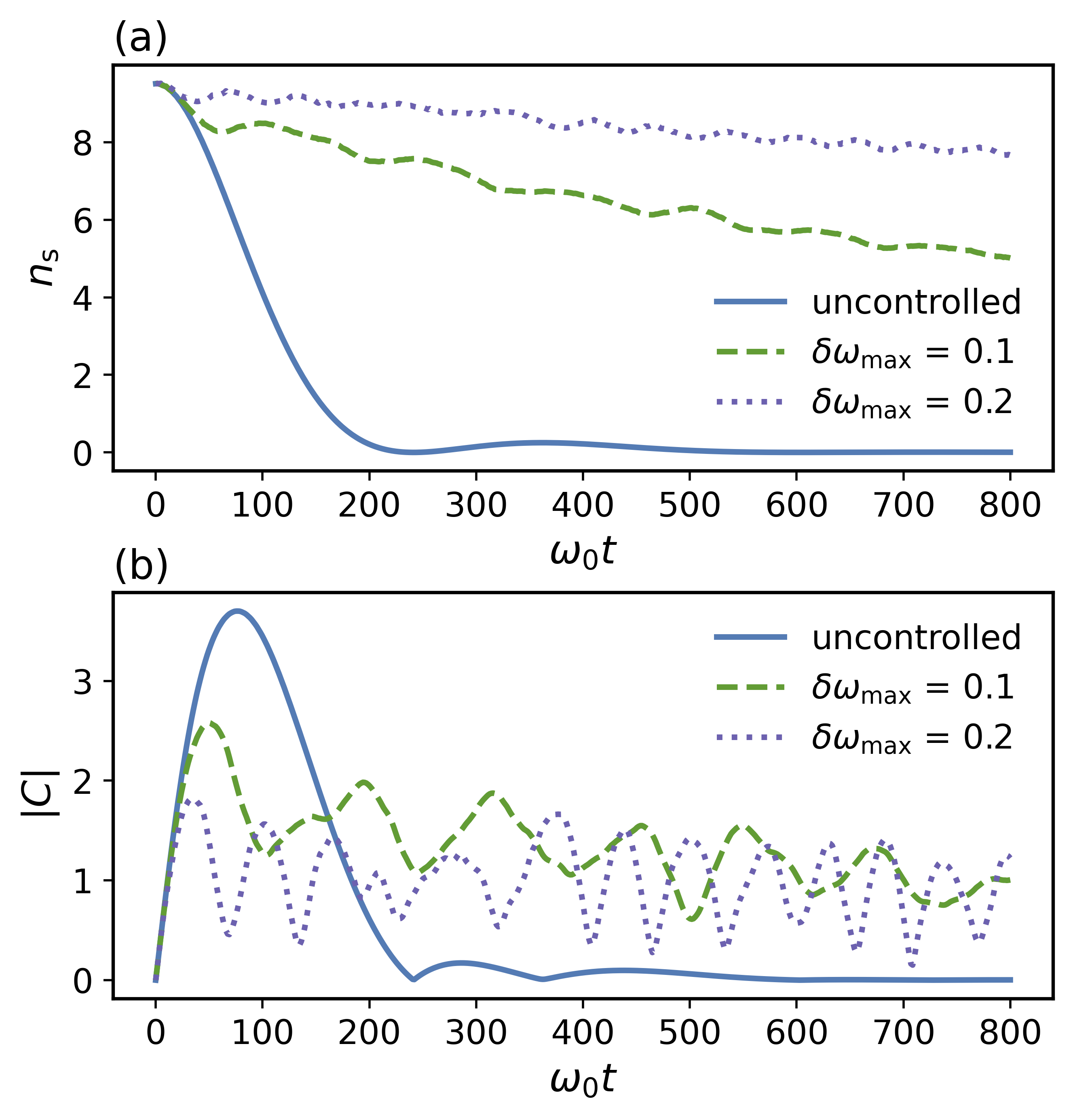}
	\caption{Effect of the maximum frequency shift $\delta\omega_\mathrm{max}$ in the random modulation on the first-order system--bath coherence dynamics. (a) System occupation number $n_\mathrm{s}$. (b) First-order system--bath coherence amplitude $|C|$. The switching interval is $\omega_0\tau_\mathrm{ran}=1.6$.}
	\label{fig7}
\end{figure}
Since random modulation gives the strongest protection in Fig.~\ref{fig6}, we further examine how its effectiveness depends on the modulation features, namely the maximum frequency shift and its switching interval. Fig.~\ref{fig7} shows the controlled dynamics of the system occupation number $n_\mathrm{s}$ and coherence amplitude $|C|$ for different maximum frequency shifts $\delta\omega_{\max}$, with the switching interval fixed at $\omega_0\tau_\mathrm{ran}=1.6$. As shown in Fig.~\ref{fig7}(a), increasing $\delta\omega_{\max}$ leads to stronger retention of the system occupation number. This indicates that a larger random frequency shift more effectively suppresses excitation leakage. The coherence dynamics in Fig.~\ref{fig7}(b) show that random modulation reduces the first coherence peak observed in the uncontrolled case, but prevents the subsequent coherence decay to zero. For the larger modulation range, $|C(t)|$ still remains finite throughout the evolution and fluctuates more strongly at long times.

\begin{figure}[!tb]
	\centering
	\includegraphics[width=\columnwidth]{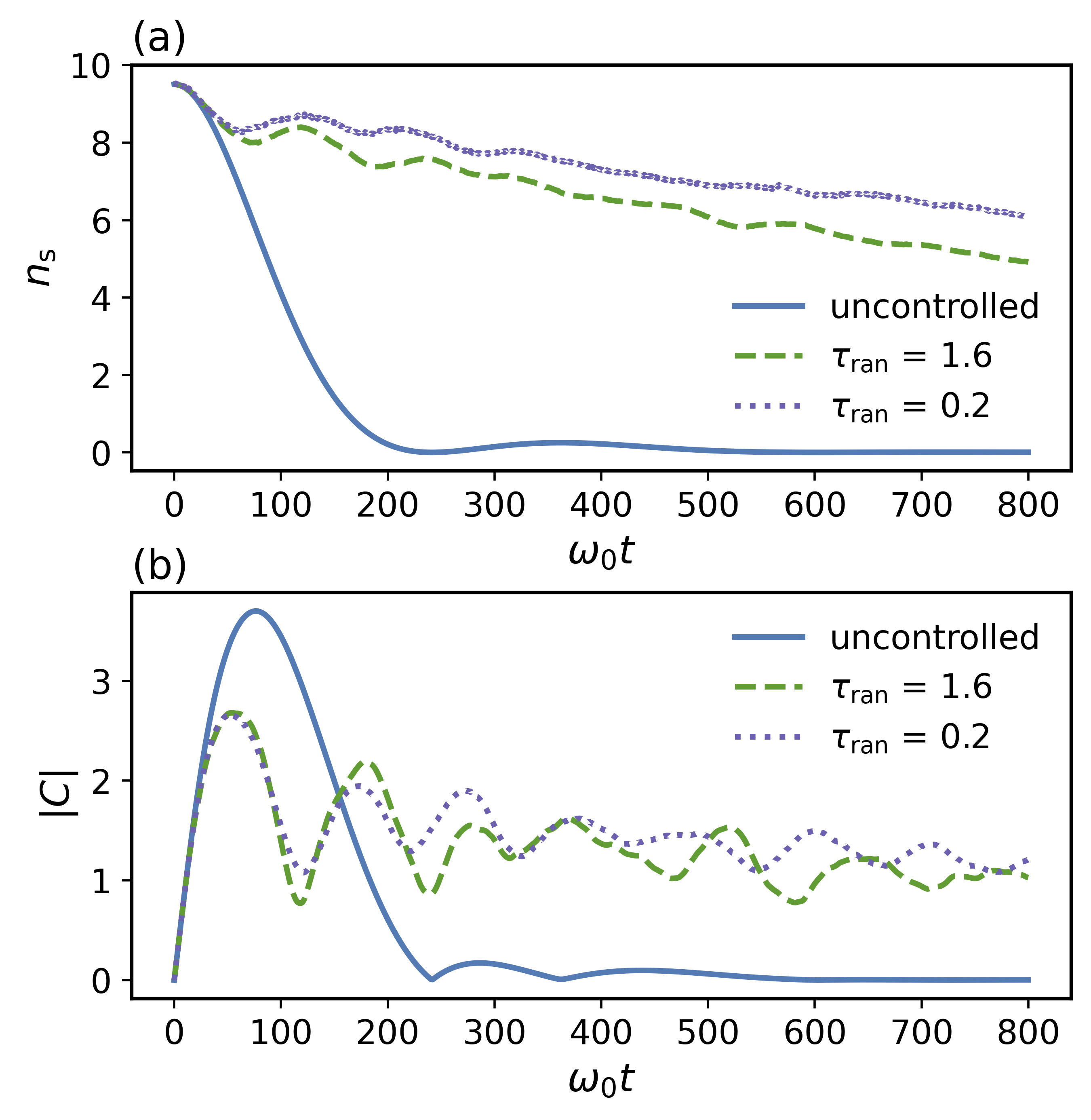}
	\caption{Effect of the switching interval $\tau_\mathrm{ran}$ in the random modulation on the first-order system--bath coherence dynamics. (a) System occupation number $n_\mathrm{s}$. (b) First-order system--bath coherence amplitude $|C|$. The maximum frequency shift is $\delta\omega_\mathrm{max}=0.1$.}
	\label{fig8}
\end{figure}
Figure~\ref{fig8} illustrates the effect of the switching interval $\tau_\mathrm{ran}$ in the random modulation, with the maximum frequency shift fixed at $\delta\omega_\mathrm{max}=0.1$. As shown in Fig.~\ref{fig8}(a), a smaller $\tau_\mathrm{ran}$ gives stronger protection of the system occupation number. For $\tau_\mathrm{ran}=0.2$, $n_\mathrm{s}(t)$ remains higher throughout the evolution, whereas for $\tau_\mathrm{ran}=1.6$, the occupation is still much better preserved than in the uncontrolled case but shows a more visible decrease at long times. This indicates that more frequent random switching more effectively suppresses the leakage of system excitation into the bath. The coherence dynamics in Fig.~\ref{fig8}(b) show a similar tendency. In the uncontrolled case, $|C(t)|$ reaches a large peak and then decays to nearly zero. Under random modulation, the peak is reduced, but $|C(t)|$ remains finite and oscillatory at long times. Compared with $\tau_\mathrm{ran}=1.6$, the smaller switching interval $\tau_\mathrm{ran}=0.2$ keeps the long-time coherence fluctuating more stably and at a slightly higher level. In brief, a shorter switching interval suppresses excitation leakage more efficiently and maintains finite and stable first-order system--bath coherence at long times.

\section{Conclusions} \label{sec_con}

In this work, we present an exact characterization of first-order system--bath coherence in bilinear bosonic models. In such models, the linearity of the Heisenberg equations allows both the system and bath operators to be expressed in terms of their initial operators, leading to closed equations for the expansion coefficients and further enabling analytical solutions for Lorentzian spectral densities. On this basis, we investigate the first-order system--bath coherence induced by excitation exchange between the system oscillator and the collective bath mode coupled to it. This coherence is obtained directly from the exact operator expansions and the initial mode occupations, without reconstructing the full joint system--bath density matrix.

We then examine the dependence of the first-order system--bath coherence on the system and bath parameters. The main results are summarized as follows. 
Increasing the system temperature raises the initial system excitation and thus enhances the transient coherence amplitude through the system-induced contribution. 
A higher bath temperature mainly affects the bath-induced contribution and thermal backflow, but only weakly modifies the total coherence amplitude in the present regime. 
The coupling strength controls the timescale of excitation exchange, and its ratio to the spectral width determines the damping character of the coherence dynamics, with stronger coupling driving the dynamics from overdamped or critically damped relaxation to underdamped oscillations. 
The spectral width sets the bath memory time, so that a narrower spectral density delays the coherence buildup but enhances its peak and leads to stronger oscillatory behavior.
Finally, detuning weakens the resonant excitation exchange and thus reduces the coherence peak while introducing faster phase-induced oscillations.

We further investigate the controlled coherence dynamics under LEO-inspired frequency modulation of the system oscillator. Two protocols are considered: \textit{rectangular modulation}, in which the system frequency shift $\delta\omega$ periodically switches between a given maximum $\delta\omega_\mathrm{max}$ and zero, and \textit{random modulation}, in which $\delta\omega$ is independently sampled from a uniform distribution over $[0, \delta\omega_\mathrm{max}]$ after each switching interval $\tau_\mathrm{ran}$. The results show that such modulation suppresses the leakage of system excitation into the bath and modifies the dynamics of the first-order system--bath coherence. Random modulation is more effective, and its protective effect is enhanced by a larger maximum frequency shift $\delta\omega_\mathrm{max}$ and a shorter switching interval $\tau_\mathrm{ran}$. In the controlled cases, the first peak of the coherence amplitude associated with rapid excitation leakage is generally reduced, but the modulation prevents the coherence amplitude from decaying to nearly zero and maintains finite and relatively stable fluctuations at long times by inhibiting complete excitation dissipation.

Bilinear bosonic models are directly connected to physical platforms where a localized bosonic mode is linearly coupled to a continuum of environmental modes. A representative example is a photonic-crystal cavity coupled to waveguides, where localized cavity modes interact with propagating waveguide modes~\cite{coles2014waveguidecoupled}. More broadly, engineered photonic- or phononic-crystal structures can produce non-flat spectral densities through their band structures, providing a physical motivation for structured bosonic reservoirs~\cite{ren2020two}. The Lorentzian spectral density used in this work should be understood as an analytically tractable effective model that captures finite bandwidth and finite memory time, rather than as a detailed microscopic spectral density of a specific platform. Our results show that first-order system--bath coherence in structured bosonic environments can be shaped both by reservoir spectral features and by external frequency modulation. Future work may extend the present framework to nonclassical initial states, counter-rotating interactions, multi-mode systems, non-equilibrium steady states, and more realistic spectral densities.

\section*{Acknowledgements}
This work is supported by the Natural Science Foundation of Shandong Province (Grant No. ZR2024MA046); the Fundamental Research Funds for the Central Universities (Grant No. 202364008); MCIN/AEI/10.13039/501100011033 (Grant No. PID2021-126273NB-I00); the European Regional Development Fund (``A way of making Europe"); the Basque Government (Grant No. IT1470-22); the Ministry for Digital Transformation and Civil Service of the Spanish Government through the Quantum Spain project (QUANTUM ENIA call); and the European Union through the Recovery, Transformation and Resilience Plan – NextGenerationEU, under the Digital Spain 2026 Agenda. Y.-Y. Xie acknowledges additional support from China Scholarship Council (No. 202406330066). P. Brumer's work was supported by the Natural Sciences and Engineering Research Council of Canada (NSERC).

\appendix
\section*{Appendix}
\setcounter{equation}{0}
\renewcommand{\theequation}{A\arabic{equation}}

In the Appendix, we consider two simple examples to clarify the distinction bewteen entanglement induced by bilinear interactions (EIBIs) and entanglement induced by nonlinear interactions (EINIs).

First, consider an ideal 50:50 beam splitter described by the unitary
\begin{align*}
	U = \exp \left[ -i\frac{\pi}4 \left( a_1^\dagger a_2 + a_1 a_2^\dagger \right) \right],
\end{align*}
which acts on creation operators of the two modes as
\begin{align*}
	U a_1^{\dagger}U^{\dagger} & = \frac{1}{\sqrt{2}} \left( a_{1}^{\dagger}-ia_{2}^{\dagger} \right), \\
	U a_2^{\dagger}U^{\dagger} & = \frac{1}{\sqrt{2}} \left( a_{2}^{\dagger}-ia_{1}^{\dagger} \right).
\end{align*}
Acting on the two-boson input state
\begin{align*}
	|1\rangle_1|1\rangle_2=a_1^\dagger a_2^\dagger|0\rangle_1|0\rangle_2,
\end{align*}
this unitary yields a nonseparable output state of the two bosonic modes,
\begin{align}
	\frac{1}{\sqrt{2}} \left(|2\rangle_1|0\rangle_2+|0\rangle_1|2\rangle_2 \right).
	\label{output1}
\end{align}
This nonseparability originates from the coherent superposition of indistinguishable bosonic paths under fixed linear mode mixing
\begin{align}
	a_i^\dagger \to \sum_j U_{ij}a_j^\dagger.
\end{align}

By contrast, consider a controlled-NOT (CNOT) gate, which acts on two logical qubits as
\begin{align*}
	|00\rangle\to|00\rangle, \quad |01\rangle\to|01\rangle,\\ 
	|10\rangle\to|11\rangle, \quad |11\rangle\to|10\rangle.
\end{align*}
The CNOT gate is locally equivalent to a controlled-phase (CPHASE) gate via single-qubit Hadamard operations~\cite{nielsen2001quantum} and can therefore be implemented using nonlinear interactions such as Kerr-type nonlinearities~\cite{scala2024deterministic}.
Under the dual-rail encoding, a logical qubit is represented as 
\begin{align*}
	|0\rangle_\mathrm{L} = |1\rangle_1 |0\rangle_2, \quad 
	|1\rangle_\mathrm{L} = |0\rangle_1 |1\rangle_2,
\end{align*}
so that the two-qubit basis states are
\begin{align*}
	|00\rangle_{\mathrm{L}} = |1\rangle_{1}|0\rangle_{2}|1\rangle_{3}|0\rangle_{4}, \\
	|01\rangle_{\mathrm{L}} = |1\rangle_{1}|0\rangle_{2}|0\rangle_{3}|1\rangle_{4}, \\
	|10\rangle_{\mathrm{L}} = |0\rangle_{1}|1\rangle_{2}|1\rangle_{3}|0\rangle_{4}, \\
	|11\rangle_{\mathrm{L}} = |0\rangle_{1}|1\rangle_{2}|0\rangle_{3}|1\rangle_{4}.
\end{align*}
For an initial product state
\begin{align*}
	\frac{|0\rangle_{\mathrm L}+|1\rangle_{\mathrm L}}{\sqrt{2}}\otimes|0\rangle_{\mathrm L} = \frac{1}{\sqrt{2}} \left( a_1^\dagger a_3^\dagger + a_2^\dagger a_3^\dagger \right) |0\rangle_1|0\rangle_2|0\rangle_3|0\rangle_4,
\end{align*}
the CNOT operation yields
\begin{align}
	\frac{|00\rangle_{\mathrm L}+|11\rangle_{\mathrm L}}{\sqrt{2}} = \frac{1}{\sqrt{2}} \left( a_{1}^{\dagger}a_{3}^{\dagger} + a_{2}^{\dagger}a_{4}^{\dagger} \right) |0\rangle_{1}|0\rangle_{2}|0\rangle_{3}|0\rangle_{4},
	\label{output2}
\end{align}
which is entangled between the two logical qubits. 
In this case, the target modes are transformed conditionally on the occupation of control modes: 
\begin{align}
	a_1^\dagger a_3^\dagger \to a_1^\dagger a_3^\dagger, \\ 
	a_2^\dagger a_3^\dagger \to a_2^\dagger a_4^\dagger. 
\end{align}

In summary, although both output states~(\ref{output1}) and (\ref{output2}) are entangled, the first arises from coherent mode mixing and can be described in terms of excitation-exchange coherence between different modes, whereas the second arises from a state-dependent transformation and cannot be characterized by such coherence. Bilinear interactions are computationally more limited because they must be supplemented with additional resources, such as ancillary states, measurements and postselection, to support universal quantum computation	~\cite{knill2001a, wu2013no}. By contrast, nonlinear interactions can directly implement deterministic conditional entangling gates, such as CNOT and CPHASE gates, which are required for universal quantum computation.

\bibliography{refs.bib}
\end{document}